\newif\ifSHOWPROOFS
\documentclass[a4paper]{llncs}
\SHOWPROOFSfalse
\SHOWPROOFStrue

\usepackage{a4,a4wide}
\usepackage{xspace}
\usepackage{times}
\usepackage{url}
\usepackage{amsmath,amssymb}
\usepackage{algorithm}
\usepackage{algorithmic}
\usepackage{graphicx}
\usepackage{pst-tree}
\usepackage{pst-node}
\usepackage{epstopdf}
\usepackage{longtable}

\newcommand{\naf}{\ensuremath{\mathrm{not}}\xspace}

\newcommand{\atom}[2]{\ensuremath{\mathtt{#1}{\rm(}\mathtt{#2}{\rm)}\xspace}}
\renewcommand{\atom}[2]{\ensuremath{\mathit{#1}{(}{#2}{)}\xspace}}

\newcommand{\gringo}{\texttt{Gringo}\xspace}
\newcommand{\aspviz}{\texttt{ASPVIZ}\xspace}
\newcommand{\idpdraw}{\texttt{IDPDraw}\xspace}

\newcommand{\dlv}{\texttt{DLV}\xspace}

\newcommand{\sealion}{\texttt{SeaLion}\xspace}
\newcommand{\kara}{\texttt{Kara}\xspace}

\newcommand{\alphabet}{\ensuremath{\mathcal{A}}\xspace}

\newcommand{\vispred}{\ensuremath{\mathcal{P}_v}\xspace}
\newcommand{\abdpred}{\ensuremath{\mathcal{P}_a}\xspace}
\newcommand{\intpred}{\ensuremath{\mathcal{P}_i}\xspace}
\newcommand{\abddom}{\ensuremath{\mathcal{D}_a}\xspace}

\newcommand{\be}{\begin{compactenum}}
\newcommand{\ee}{\end{compactenum}}
\newcommand{\bi}{\begin{compactitem}}
\newcommand{\ei}{\end{compactitem}}

\newcommand{\iec}[0]{i.e.,\xspace}

\newcommand{\egc}[0]{e.g.,\xspace}

\newcommand{\nop}[1]{}
\newcommand{\body}[1]{\ensuremath{\mathrm{B(}{#1}\mathrm{)}}\xspace}
\newcommand{\head}[1]{\ensuremath{\mathrm{H(}{#1}\mathrm{)}}\xspace}
\newcommand{\pbody}[1]{\ensuremath{\mathrm{{B^+}(}{#1}\mathrm{)}}\xspace}

\newcommand{\trans}[2]{\ensuremath{\lambda{\rm(}{#1},{#2}{\rm)}}}
\newcommand{\guesspart}[1]{\ensuremath{\mathit{guess}{\rm(}{#1}{\rm)}}}
\renewcommand{\guesspart}[1]{\ensuremath{\mathsf{guess}{\rm(}{#1}{\rm)}}}
\newcommand{\checkpart}[1]{\ensuremath{\mathit{check}{\rm(}{#1}{\rm)}}}
\renewcommand{\checkpart}[1]{\ensuremath{\mathsf{check}{\rm(}{#1}{\rm)}}}
\newcommand{\var}[1]{\ensuremath{\mathit{VAR}{\rm(}{#1}{\rm)}}}
\newcommand{\dompart}[2]{\ensuremath{\mathit{dom}{\rm(}{#1},{#2}{\rm)}}}
\renewcommand{\dompart}[2]{\ensuremath{\mathsf{dom}{\rm(}{#1},{#2}{\rm)}}}

\newcommand{\ruleimp}{\mathtt{{:-}}}
\renewcommand{\ruleimp}{:\!- \ }

\newcommand{\nonrecdomo}[0]{\ensuremath{\mathtt{nonRecDom}}}
\renewcommand{\nonrecdomo}[0]{\ensuremath{\mathit{nonRecDom}}}
\newcommand{\nonrecdom}[1]{\ensuremath{\nonrecdomo(#1)}}
\newcommand{\domo}[0]{\ensuremath{\mathtt{dom}}}
\renewcommand{\domo}[0]{\ensuremath{\mathit{dom}}}
\newcommand{\dom}[1]{\ensuremath{\domo(#1)}}

\title{\kara: A System for Visualising and Visual Editing of Interpretations for Answer-Set Programs%
\thanks{This work was partially supported by the Austrian Science Fund (FWF)
under project P21698.}}

\author{%
Christian Kloim\"ullner\inst{1} \and Johannes Oetsch\inst{2} \and J\"org P\"uhrer\inst{2} \and Hans Tompits\inst{2}}

\institute{%
Forschungsgruppe f\"ur Industrielle Software (INSO),\\
Technische Universit\"at Wien,\\
Favoritenstra\ss{}e\ 9-11,
A-1040 Vienna, Austria \\
\email{christian.kloimuellner@inso.tuwien.ac.at}
\and
Institut f\"ur Informationssysteme 184/3,\\
Technische Universit\"at Wien,\\
Favoritenstra\ss{}e\ 9-11,
A-1040 Vienna, Austria \\
\email{\{oetsch,puehrer,tompits\}@kr.tuwien.ac.at}
}

\begin{document}

\maketitle
                                              
\begin{abstract}
In answer-set programming (ASP), the solutions of a problem
are encoded in dedicated models, called \emph{answer sets}, of a logical theory.
These answer sets are computed  from the program that represents the theory by
means of an ASP solver and returned to the user as sets of ground first-order literals.
As this type of representation is often cumbersome for the user to interpret,
tools like \aspviz and \idpdraw were developed that allow for
visualising answer sets.
The tool \kara, introduced in this paper, follows these approaches, using ASP itself as a language for defining
visualisations of interpretations.
Unlike existing tools 
that position graphic primitives according to static coordinates only,
\kara allows for more high-level specifications, supporting
graph structures, grids, and relative positioning of graphical elements.
Moreover, generalising the functionality of previous tools, \kara provides modifiable visualisations such that
interpretations can be manipulated by graphically editing their visualisations.
This is realised by resorting to abductive reasoning techniques.
\kara is part of \sealion, a forthcoming integrated development environment (IDE) for ASP.

\end{abstract}
\section{Introduction}
Answer-set programming (ASP)~\cite{baral03} is a well-known paradigm for declarative problem solving.
Its key idea is that a problem is encoded in terms of a logic program such that dedicated models
of it, called \emph{answer sets}, correspond to the solutions of the problem.
Answer sets are interpretations, usually represented by  sets of  ground first-order literals.

A problem often faced when developing answer-set programs is that interpretations returned by an ASP solver are cumbersome to read---in particular, in case of large interpretations which are spread over several lines on the screen or the output file.
Hence, a user may have difficulties extracting the information he or she is
interested in from the textual representation of an answer set.
Related to this issue, there is one even harder practical problem: editing or writing interpretations
by hand.

Although the general goal of ASP is to have answer sets computed automatically,
we identify different situations during the development of answer-set programs in which it would be helpful
to have adequate means to manipulate interpretations.
First, in declarative debugging~\cite{shapiro1982}, the user has to specify the semantics he or she expects in order for
the debugging system to identify the causes for a mismatch with the actual semantics.
In previous work~\cite{Oetsch-etal09-debug}, a debugging approach has been introduced
that takes a program $P$ and an interpretation $I$
that is expected to be an answer set of $P$ and returns reasons why $I$ is not an answer set of $P$.
Manually producing such an intended interpretation ahead of computation is a time-consuming task, however.
Another situation in which the creation of an interpretation 
can be useful is testing post-processing tools.
Typically, if answer-set solvers are used within an online application, they are embedded as a module in
a larger context.
The overall application delegates a problem to the solver by transforming it to a respective
answer-set program and the outcome of the solver is then 
processed further as needed by the application.
In order to test post-processing components, which may be written by programmers unaware of ASP,
it would be beneficial to have means to create mock answer sets as test inputs.
Third, the same idea of providing test input applies to modular answer-set programming~\cite{JanhunenOTW09},
when a module $B$ that depends on another module $A$ is developed before or separately from $A$.
In order to test $B$, it can be joined with interpretations mocking answer sets from $A$.

In this paper, we describe the system \kara which allows for both
visualising interpretations and editing them by manipulating their visualisations.\footnote{The name ``\kara'' derives, with all due respect, from ``Kara Zor-El'', the native Kryptonian name of \emph{Supergirl}, given that Kryptonians have visual superpowers on Earth.}
The visualisation functionality of \kara 
has been inspired by the existing tools \aspviz~\cite{CliffeVBP08} and \idpdraw~\cite{idpdraw} for visualising answer sets.
The key idea is to use ASP itself as a language for specifying
how to visualise an interpretation $I$.
To this end, the user takes a dedicated answer-set program $V$---which we call a \emph{visualisation program}---that specifies how the visualisation of $I$ should look like.
That is, $V$ defines how different {graphical elements}, such as rectangles, polygons, images, graphs, etc., should be
arranged and configured to visually represent $I$.

\kara offers a rich visualisation language
that allows for defining a superset of the graphical elements available in \aspviz and \idpdraw,
e.g., providing support for automatically layouting graph structures,
relative and absolute positioning,
and support for grids of graphical elements.
Moreover, \kara also offers a \emph{generic mode} of visualisation, not available in previous tools,
that does not require a domain-specific visualisation program,
representing an answer set as a hypergraph whose set of nodes corresponds to the individuals
occurring in the interpretation.\footnote{%
A detailed overview of the differences concerning the visualisation capabilities of \kara
with other tools is given in Section~\ref{sec:related}.}
A general difference to previous tools is that \kara does not just produce image files right away but
presents the visualisation in form of modifiable graphical elements in a visual editor.
The user can manipulate the visualisation in various ways,
e.g., change size, position, or other properties of graphical elements,
as well as copy, delete, and insert new graphical elements.
Notably, the created visualisations can also be used outside
our editing framework, as \kara offers an SVG export function that allows to save the
possibly modified visualisation as a vector graphic.
Besides fine-tuning exported SVG files, 
manipulation of the visualisation of an interpretation $I$
can be done for obtaining a modified version $I'$ of $I$
by means of abductive reasoning~\cite{Peirce55abductionand}.
This gives the possibility to visually edit interpretations
which is useful for debugging and testing purposes as described above.

In Section~\ref{sec:example}, we present a number of examples
that illustrate the functionality of \kara and the ease of coping with
a visualised answer set compared to interpreting its textual representation.

\kara is designed as a plugin of \sealion, an Eclipse-based integrated development environment (IDE) for 
ASP~\cite{sealion}
that is currently developed as part of a project on programming-support methods for ASP~\cite{mmdasp}.

\begin{figure}[t]
\centering
  \includegraphics[scale=0.8]{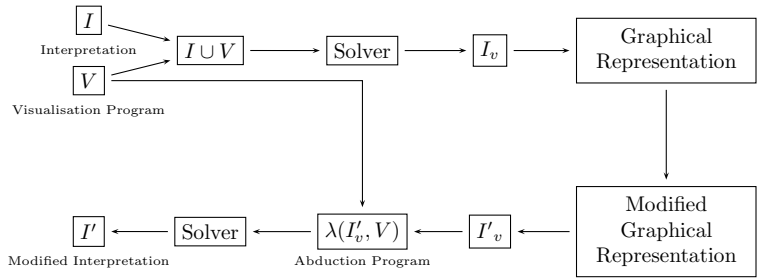}
  \caption{Overview of the workflow (visualisation and abduction process).}
\label{fig:workflow}
\end{figure}

\section{System Overview}\label{sec:overview}
We assume familiarity with the basic concepts of answer-set programming (ASP) (for a thorough introduction to the subject, cf.\ Baral~\cite{baral03}).
In brief, an answer-set program consists of rules of the form
\[
a_1 \vee \cdots \vee a_l \ruleimp a_{l+1}, \ldots, a_{m}, \naf\ a_{m+1}, \ldots, \naf\ a_{n} , 
\]
where $n \geq m \geq l \geq 0$, ``$\naf$'' denotes \emph{default negation}, and  all $a_i$ are first-order literals (\iec atoms possibly preceded by the \emph{strong negation} symbol, $\neg$).
For a rule $r$ as above, we define the \emph{head}  of $r$ as $\head{r} = \{a_1,  \ldots,  a_l\}$ and
the \emph{positive body} as $\pbody{r} = \{a_{l+1}, \ldots, a_{m}\}$. 
If $n=l=1$,
$r$  is a \emph{fact}, 
and
if $l=0$,
$r$ is a \emph{constraint}. For facts, we will usually omit the symbol ``$\ruleimp$''.
The \emph{grounding} of a  program $P$ relative to its Herbrand universe 
is defined as usual.
An \emph{interpretation} $I$ is a finite and consistent set of ground literals, where consistency means that $\{a,\neg a\}\not\subseteq I$, for any atom $a$.
$I$ is an \emph{answer set} of a program $P$ if it is a minimal model of the grounding of the \emph{reduct} of $P$ relative to $I$ (see Baral~\cite{baral03} for details).

The overall workflow of \kara is depicted in
Fig.~\ref{fig:workflow}, illustrating how an interpretation $I$
can be visualised in the upper row and how changing the visualisation
can be reflected back to $I$ such that we obtain a modified version $I'$ of $I$ in the lower row.
In the following, we call programs that encode problems 
for which
$I$ and $I'$ provide solution candidates \emph{domain programs}.

\subsection{Visualisation of Interpretations}
As discussed in the introduction, we use ASP itself as a language for specifying
how to visualise an interpretation.
In doing so, we follow a similar approach as the tools  \aspviz~\cite{CliffeVBP08} and \idpdraw~\cite{idpdraw}.
We next describe  this method on an abstract level.

Assume we want to visualise an interpretation $I$ that is defined over a first-order alphabet \alphabet.
We  join $I$, interpreted as a set of facts, with a visualisation program $V$
that is defined over $\alphabet'\supset\alphabet$, where $\alphabet'$ may contain auxiliary predicates and function symbols,
as well as predicates from a fixed set \vispred of reserved \emph{visualisation predicates} that vary for the three tools.%
\footnote{Technically, in \aspviz, $V$ is not  joined with $I$ but with a domain program $P$ such that $I$ is an answer set of $P$.}

The rules in $V$ are used to derive different atoms with predicates from \vispred, depending on $I$,
that control the individual graphical elements of the resulting visualisation
including their presence or absence, position, and all other properties.
An actual visualisation is obtained by post-processing an answer set $I_v$ of $V\cup I$ that is projected to the predicates in \vispred.
We refer to $I_v$ as a \emph{visualisation answer set} for $I$.
The process is depicted in the upper row of Fig.~\ref{fig:workflow}.
An exhaustive list of visualisation predicates available in \kara is given in Appendix~\ref{app:vispred}.

\begin{example}\label{ex:simple}
Assume we deal with a domain program whose answer sets correspond to arrangements of items on two shelves.
Consider the interpretation 
$I=\{\atom{book}{s_1, 1},$ $\atom{book}{s_1, 3},\atom{book}{s_2,1},$ $\atom{globe}{s_2,2}\}$
stating that two books are located on shelf $s_1$ in positions $1$ and $3$
and that there is another book and a globe on shelf $s_2$ in positions $1$ and $2$.
% 
%%%%%%%%%%%%%%%%%%%%%%%%%%%%%%%%%%%%%%%
% 
\begin{figure}[t]
\centering
\includegraphics[scale=1.0]{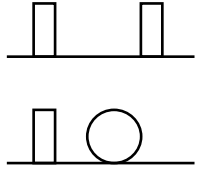}
\caption{The visualisation of interpretation $I$ from Example~\ref{ex:simple}.}
\label{simpleimg}
\end{figure}
%
%%%%%%%%%%%%%%%%%%%%%%%%%%%%%%%%%%%%%%%%%%%%%%
% 
The goal is to create a simple graphical representation of this and similar interpretations,
depicting the two shelves as two lines,
each book as a rectangle, and globes as circles.
Consider the following visualisation program:
\begin{eqnarray}
% \mbox{\emph{\%}} & \mbox{\emph{Rule 1-2: static lines representing shelfs.}} \\
&&  \mathit{visline}(\mathit{shelf}_1,10,40,80,40,0).\label{eq:ex1:1}\\[-.7ex]
&&  \mathit{visline}(\mathit{shelf}_2,10,80,80,80,0).\label{eq:ex1:2}\\%[.2ex]
% \mbox{\emph{\%}} & \mbox{\emph{Rule 3-5: display books.}}\\
&&  \mathit{visrect}(f(X,Y),20,8) \ruleimp \mathit{book}(X,Y).\label{eq:ex1:3}\\[-.7ex]
&&  \mathit{visposition}(f(s_1,Y),20*Y,20,0) \ruleimp \mathit{book}(s_1,Y).\label{eq:ex1:4}\\[-.7ex]
&&  \mathit{visposition}(f(s_2,Y),20*Y,60,0) \ruleimp \mathit{book}(s_2,Y).\label{eq:ex1:5}\\%[.2ex]
% \mbox{\emph{\%}} & \mbox{\emph{Rule 6-8: display globes.}}\\
&&  \mathit{visellipse}(f(X,Y),20,20) \ruleimp \mathit{globe}(X,Y).\label{eq:ex1:6}\\[-.7ex]
&&  \mathit{visposition}(f(s_1,Y),20* Y,20,0) \ruleimp \mathit{globe}(s_1,Y).\label{eq:ex1:7}\\[-.7ex]
% \end{array}$$
% $$
% \begin{array}{ll}
&&  \mathit{visposition}(f(s_2,Y),20* Y,60,0) \ruleimp \mathit{globe}(s2,Y).\label{eq:ex1:8}
\end{eqnarray}

% $$
% \begin{array}{ll}
% \mbox{\emph{\%}} & \mbox{\emph{Rule 1-2: static lines representing shelfs.}} \\
% &  \mathit{visline}(\mathit{shelf}_1,10,40,80,40,0).\quad
%   \mathit{visline}(\mathit{shelf}_2,10,80,80,80,0).\\[.2ex]
% \mbox{\emph{\%}} & \mbox{\emph{Rule 3-5: display books.}}\\
% &  \mathit{visrect}(f(X,Y),20,8) \ruleimp \mathit{book}(X,Y).\\
% &  \mathit{visposition}(f(s_1,Y),20*Y,20,0) \ruleimp \mathit{book}(s_1,Y).\\
% &  \mathit{visposition}(f(s_2,Y),20*Y,60,0) \ruleimp \mathit{book}(s_2,Y).\\[.2ex]
% \mbox{\emph{\%}} & \mbox{\emph{Rule 6-8: display globes.}}\\
% &  \mathit{visellipse}(f(X,Y),20,20) \ruleimp \mathit{globe}(X,Y).\\
% &  \mathit{visposition}(f(s_1,Y),20* Y,20,0) \ruleimp \mathit{globe}(s_1,Y).\\
% % \end{array}$$
% % $$
% % \begin{array}{ll}
% &  \mathit{visposition}(f(s_2,Y),20* Y,60,0) \ruleimp \mathit{globe}(s2,Y).
% \end{array}
% $$
% 

\noindent
Rules~(\ref{eq:ex1:1}) and (\ref{eq:ex1:2}) create two lines with the identifiers $\mathit{shelf}_1$ and $\mathit{shelf}_2$,
representing the top and bottom shelf.
The second to fifth arguments of $\mathit{visline}/6$ represent the origin and the target coordinates of the line.\footnote{The origin of the coordinate system is at the top-left corner of the illustration window with the 
$x$-axis pointing to the right and the $y$-axis pointing down.}
The last argument of $\mathit{visline}/6$ is a $z$-coordinate determining which graphical element is visible in case two or more overlap.
Rule~(\ref{eq:ex1:3}) generates the rectangles representing books, 
and Rules~(\ref{eq:ex1:4}) and (\ref{eq:ex1:5}) determine their position depending on the shelf and the position given in the interpretation.
Likewise, Rules~(\ref{eq:ex1:6}) to (\ref{eq:ex1:8}) generate and position globes.
The resulting visualisation of $I$ is depicted in Fig.~\ref{simpleimg}.
\qed\end{example}

Note that the first argument of each visualisation predicate is a unique identifier for the respective
graphical element.
By making use of function symbols with variables, like $f(X,Y)$ in Rule~(\ref{eq:ex1:3}) above,
these labels are not limited to constants in the visualisation program
but can be generated on the fly, depending on the interpretation to visualise.
While some visualisation predicates,
like $\mathit{visline}$, $\mathit{visrect}$, and $\mathit{visellipse}$,
define graphical elements,
others, \egc $\mathit{visposition}$, are used to change properties of the elements,
referring to them by their respective identifiers.

\kara also offers a \emph{generic visualisation} that 
visualises an arbitrary interpretation without the need for defining a visualisation program.
In such a case, the interpretation is represented as a labelled hypergraph.
Its nodes are the individuals appearing in the interpretation and the edges represent the literals in the interpretation, connecting the individuals appearing in the respective literal.
Integer labels on the endings of the edge are used for expressing the term position of the individual.
To distinguish between different predicates, each edge has an additional label stating the predicate.
Edges of the same predicate are of the same colour.
A generic visualisation is presented in Example~\ref{ex:generic}. 

\subsection{Editing of Interpretations}
We next describe how we can 
obtain a modified version $I'$ of an interpretation $I$ corresponding to a manipulation of the visualisation of $I$. We follow the steps depicted in the lower row of Fig.~\ref{fig:workflow}, using abductive reasoning.
Recall that abduction is the process of finding hypotheses that explain given observations in the context of a theory.
Intuitively, in our case, the theory is the visualisation program, the observation is the modified visualisation of $I$, and the desired hypothesis is $I'$.

In \kara, the visualisation of $I$ is created using the Graphical Editing Framework (GEF)~\cite{gef} of Eclipse.
It is displayed in a graphical editor which allows for various kinds of manipulation actions
such as 
moving, resizing, adding or deleting graphical elements,
adding or removing edges between them,
editing their properties, or
change grid values.
Each change in the visual editor of \kara is internally reflected by a modification to 
the underlying visualisation answer set $I_v$.
We denote the resulting visualisation interpretation by $I'_v$.
From that and the visualisation program $V$, we construct
a logic program $\trans{I'_v}{V}$ such that the visualisation of any 
answer set  $I'$ of  $\trans{I'_v}{V}$ using $V$
corresponds to the modified one.

The idea is that $\trans{I'_v}{V}$, which we refer to as the \emph{abduction program}
for $I'_v$ and $V$, guesses a set of \emph{abducible atoms}. 
On top of these atoms, the rules of $V$ are used in $\trans{I'_v}{V}$ to derive a hypothetical
visualisation answer set $I''_v$ for $I'$.
Finally, constraints in the abduction program ensure that $I''_v$ coincides with
the targeted visualisation interpretation $I'_v$ on a set \intpred of selected predicates from \vispred,
which we call \emph{integrity predicates}.
Hence, a modified interpretation $I'$ can be obtained by computing an answer set of 
$\trans{I'_v}{V}$ and projecting it to the guessed atoms.
To summarise, the abduction problem underlying the described process can be stated as follows: 
\begin{description}
\item[($*$)] Given the interpretation $I'_v$, determine an interpretation $I'$ such that $I'_v$ coincides with each answer set of $V \cup I'$ on  \intpred .
\end{description}

Clearly, visualisation programs must be written in a way that manipulated visualisation interpretations
could indeed be the outcome of the visualisation program for some input.
This is not the case for arbitrary visualisation programs,  but
usually it is easy to write an appropriate visualisation program that allows for abducing interpretations.

\begin{figure}[t]
\hrule
\medskip
$$
\begin{array}{r@{~}l}
\dompart{I'_v}{V}=&\{
  \nonrecdom{t} \ruleimp v(\vec{t'})\mid
    r\in V, v/m\in\vispred,
    v(\vec{t'})\in\head{r},\\
&\qquad  a(\vec{t})\in\pbody{r},
    \vec{t}=t_1,\dots,t,\dots,t_n,
    a/n\notin\vispred,\\&
    \qquad \var{t}\neq\emptyset,\var{t}\subseteq\var{\vec{t'}}\}\cup%
\\[1ex]
% \end{array}
% $$ $$
% \begin{array}{r@{~}l}
 & \{\dom{t} \ruleimp v(\vec{t'}), \nonrecdom{X_1}, \dots, \nonrecdom{X_l} \mid r\in V,\\
   &\qquad v/m\in\vispred,
    v(\vec{t'})\in\head{r},
    a(\vec{t})\in\pbody{r},
    \vec{t}=t_1,\dots,t,\dots,t_n,\\
& \qquad a/n\notin\vispred,
    \var{t}\cap\var{\vec{t'}}\neq\emptyset,\\
& \qquad \var{t}\setminus\var{\vec{t'}}=\{X_1,\dots, X_l\}\}\cup\\[1ex]
  &\{\dom{X} \ruleimp \nonrecdom{X}\},\\[2ex]
\guesspart{V}=& \{
  a(X_1,\dots, X_n) \ruleimp \naf~\neg a(X_1,\dots, X_n), \dom{X_1},\dots,\dom{X_n},\\&
  \hphantom{\{}\neg a(X_1,\dots, X_n) \ruleimp \naf~a(X_1,\dots, X_n), \dom{X_1},\dots,\dom{X_n} \mid \\
& \qquad \quad a/n \notin \vispred,
    a(t_1,\dots,t_n)\in \bigcup_{r\in V}\body{r},\\
& \qquad \quad \{a(t'_1,\dots,t'_n)\mid a(t'_1,\dots,t'_n)\in\head{r},r\in V\}=\emptyset\},\\[2ex]
\checkpart{I'_v}=&\{\ruleimp \naf~v(t_1,\dots,t_n),\;
\ruleimp v(X_1,\dots,X_n), \naf~v'(X_1,\dots,X_n),\\&
		  \phantom{\{} v'(t_1,\dots,t_n)\mid v(t_1,\dots,t_n)\in I'_v, v/n\in\intpred\},
\end{array}
$$
\smallskip
\hrule
\caption{Elements of the abduction program $\trans{I'_v}{V}$.\label{def:abduction:program}}
 \end{figure}

The following problems have to be addressed for realising the sketched approach:
\begin{itemize}
\item
determining 
the predicates and domains of the
abducible atoms, and 
\item choosing
the integrity predicates among the visualisation predicates.
\end{itemize}
For solving these issues, we rely on pragmatic choices that seem useful in practice.
We obtain the set \abdpred of predicates of the abducible atoms from the visualisation program $V$.
The idea is that every predicate that is relevant to the solution of a problem encoded
in an answer set has to occur in the visualisation program if the latter is meant to 
provide a complete graphical representation of the solution.
Moreover, we restrict \abdpred to those non-visualisation predicates in $V$ that occur
in the body of a rule but not in any head atom in $V$.
The assumption is that atoms defined in $V$ are most likely of auxiliary nature
and not contained in a domain program.

An easy approach for generating a domain \abddom of the abducible atoms
would be to extract the terms occurring in $I'_v$.
We follow, however, a more fine-grained approach that takes
the introduction and deletion of function symbols in the rules in $V$
into account.
Assume $V$ contains the rules
\[
\begin{array}{r@{~}c@{~}l}
\mathit{visrect}(f(\mathit{Street},\mathit{Num}),9,10) &\ruleimp& \mathit{house}(\mathit{Street},\mathit{Num}) \quad \mbox{ and}\\
\mathit{visellipse}(\mathit{sun},\mathit{Width},\mathit{Height}) &\ruleimp& \mathit{property}(\mathit{sun},\mathit{size}(\mathit{Width},\mathit{Height})),
\end{array}
\]
and $I'_v$ contains  $\mathit{visrect}(f(\mathit{bakerstreet},221b),9,10)$ and $\mathit{visellipse}(\mathit{sun},$ $10,11)$.
Then, when extracting the terms in $I'_v$,
the domain includes $f(\mathit{bakerstreet},221b)$, $\mathit{bakerstreet}$, $221b$, $9$, $10$, $\mathit{sun}$, and $11$
for the two rules.
However,
the functor $f$ is solely an auxiliary concept in $V$ and not meant to be part of domain programs.
Moreover, the term $9$ is introduced in $V$ and is not needed in the domain for $I'$.
Also, the terms $10$ and $11$ as standalone terms and $\mathit{sun}$ are not needed in $I'$
to derive $I'_v$.
Even worse, the term $\mathit{size}(10, 11)$, that has to be contained in $I'$ such that
$I'_v$ can be a visualisation answer set for $I'$, is missing in the domain.
Hence, we derive \abddom in $\trans{I'_v}{V}$ not only from $I'_v$
but also consider the rules in $V$.
Using our translation that is detailed below, we obtain 
$\mathit{bakerstreet}$, $221b$, and $\mathit{size}(10, 12)$
as domain terms from the rules above.

For the choice of \intpred, \iec of the predicates on which
$I'_v$ and the actual visualisation answer sets of $I'$
need to coincide,
we exclude visualisation predicates that 
require a high preciseness in visual editing by the user 
in order to match exactly a value that could result from the visualisation program.
For example,
we do not include predicates determining position and size of graphical elements,
since in general it is hard to position and scale an element precisely
such that an interpretation $I'$ exists with a matching visualisation.
Note that this is not a major restriction,
as in general it is easy to write a visualisation program
such that aspects that the user wants to be modifiable 
are represented by graphical elements that can be elegantly modified visually.
For example, instead of representing a Sudoku puzzle by labels
whose exact position is calculated in the visualisation program,
the language of \kara allows for using a logical grid such that
the value of each cell can be easily changed in the visual editor.

We next give the details of the abduction program.
\begin{definition}
Let $I'_v$ be an interpretation with atoms over predicates in \vispred, $V$ a {\rm (}visualisation{\rm )} program,
and $\intpred\subseteq\vispred$ the fixed set of integrity predicates.
Moreover, let $\var{T}$ denote the variables occurring in $T$, where $T$ is a term or a list of terms.
Then, 
the \emph{abduction program} with respect to $I'_v$ and $V$ is given by
$$\trans{I'_v}{V}= \dompart{I'_v}{V} \cup \guesspart{V} \cup V \cup \checkpart{I'_v}\mbox{,}$$
where $\dompart{I'_v}{V}$, $\guesspart{V}$, and $\checkpart{I'_v}$ are given in Fig.~\ref{def:abduction:program},
and $\nonrecdomo/1$, $\domo/1$,  and $v'/n$, for all $v/n\in\intpred$, are fresh predicates.
\end{definition}
The idea of $\dompart{I'_v}{V}$ is to consider non-ground terms $t$ contained in the body of a visualisation rule 
that share variables with a visualisation atom in the head of the rule and to derive instances of these terms
when the corresponding visualisation atom is contained in $I'_v$.
In case less variables occur in the visualisation atom than in $t$, we avoid safety problems by restricting their scope
to parts of the derived domain.
Here, the distinction between predicates \domo\ and \nonrecdomo\ is necessary to prevent infinite groundings of the abduction program.
Note that in general it is not guaranteed that the domain we derive contains all necessary elements for abducing
an appropriate interpretation $I'$.
For instance, consider the case that the visualisation program contains a rule
$\mathit{visrect{\rm(}id,5,5{\rm)}}\ruleimp \mathit{foo{\rm(}X{\rm)}}$,
and $V$ together with the constraints in $\checkpart{I'_v}$ require that for all terms $t$  of a domain that can be obtained
from $I'_v$ and $V$, $\mathit{foo{\rm(}t{\rm)}}$ must not hold.
Then, there is no interpretation that will trigger the rule using this domain,
although an interpretation with a further term $\mathit{t'}$ might exist that results in the desired visualisation.
Hence, we added an editor to \kara that allows for changing and extending the automatically generated domain
as well as the set of abducible predicates.

The following result characterises the answer sets of the abduction program.
\begin{theorem}
Let $I'_v$ be an interpretation with atoms over predicates in \vispred,
$V$ a {\rm (}visualisation{\rm )} program, and $\intpred\subseteq\vispred$ the fixed set of integrity predicates.
Then, any answer set $I''_v$ of $\trans{I'_v}{V}$ coincides with $I'_v$ on the atoms over predicates from $\intpred$, and a solution $I'$ of the abduction problem {\rm (}$*${\rm )} 
is obtained from $I''_v$ by projection to the predicates in
$$\{a/n\mid a(t_1,\dots,t_n)\in \bigcup_{r\in V}\body{r},
 \{a(t'_1,\dots,t'_n)\mid a(t'_1,\dots,t'_n)\in\head{r},r\in V\}=\emptyset\}\setminus \vispred.$$
\end{theorem}

%%%%%%%%%%%%%%%%%%%%%%%%%%%%%%%%%%%%%%%%%
\begin{figure}[t!]
\hrule
\vspace{-1ex}
\begin{eqnarray}%{rl}
% &\mbox{\em \%} & \mbox{\em Rules 1-2.}\\
&&  \mathit{visgrid}(\mathit{maze},\mathit{MAXR},\mathit{MAXC},\mathit{MAXR}{*}20{+}5,\mathit{MAXC}{*}20{+}5) {\ruleimp} \mathit{maxC}(\mathit{MAXC}),
% \\
% &&\hphantom{\mathit{visgrid}(\mathit{maze},\mathit{MAXR},\mathit{MAXC},\mathit{MAXR}{*}20{+}5,\mathit{MAXC}{*}20{+}5) \ruleimp }                                                 
\mathit{maxR}(\mathit{MAXR}).\label{ex:2:1}\\[-.7ex]
&&  visposition(maze,0,0,0).\label{ex:2:2}\\%[1ex]
&\mbox{\tt \%} & \mbox{\tt %Rules 3-4: 
A cell with a wall on it.}\nonumber\\[-.7ex]
&&  \mathit{visrect}(\mathit{wall},20,20).\label{ex:2:3} \\[-.7ex]
&& \mathit{visbackgroundcolor}(\mathit{wall},\mathit{black}).\label{ex:2:4}\\%[1ex]
& \mbox{\tt \%} & \mbox{\tt %Rules 5-7: 
An empty cell.}\nonumber\\[-.7ex]
&&  \mathit{visrect}(\mathit{empty},20,20).\label{ex:2:5} \\[-.7ex]
&& \mathit{visbackgroundcolor}(\mathit{empty},\mathit{white}).\label{ex:2:6}\\[-.7ex]
&& \mathit{viscolor}(\mathit{empty},\mathit{white}).\label{ex:2:7}\\%[1ex]
&\mbox{\tt \%} & \mbox{\tt %Rules 8-11: e
Entrance and exit.}\nonumber\\[-.7ex]
&&  \mathit{visimage}(\mathit{entrance},``\mathit{entrance.jpg}").\label{ex:2:8} \\[-.7ex]
 && \mathit{visscale}(\mathit{entrance},18,18).\label{ex:2:9} \\[-.7ex]
&& \mathit{visimage}(\mathit{exit},``\mathit{exit.png}").\label{ex:2:10} \\[-.7ex]
&& \mathit{visscale}(\mathit{exit},18,18).\label{ex:2:11}\\%[1ex]
& \mbox{\tt \%} & \mbox{\tt %Rules 12-15: f
Filling the cells of the grid.}\nonumber\\[-.7ex]
&&  \mathit{visfillgrid}(\mathit{maze},\mathit{empty},R,C)\ruleimp \mathit{empty}(C,R), \naf\ \mathit{entrance}(C,R),
                                              \naf\ \mathit{exit}(C,R).\label{ex:2:12}\\[-.7ex]
&&  \mathit{visfillgrid}(\mathit{maze},\mathit{wall},R,C) \ruleimp \mathit{wall}(C,R),\naf\ \mathit{entrance}(C,R),
                                              \naf\ \mathit{exit}(C,R).\label{ex:2:13}\\[-.7ex]
&&  \mathit{visfillgrid}(\mathit{maze},\mathit{entrance},R,C) \ruleimp \mathit{entrance}(C,R).\label{ex:2:14}\\[-.7ex]
&&  \mathit{visfillgrid}(\mathit{maze},\mathit{exit},R,C) \ruleimp \mathit{exit}(C,R).\label{ex:2:15}\\%[1ex]
& \mbox{\tt \%} & \mbox{\tt % Rules 16-19: v
Vertical and horizontal lines.}\nonumber\\[-.7ex]
&&  \mathit{visline}(v(0),5,5,5,\mathit{MAXR}*20+5,1) \ruleimp \mathit{maxR}(\mathit{MAXR}).\label{ex:2:16}\\[-.7ex]
&&  \mathit{visline}(v(C),C{*}20{+}5,5,C{*}20{+}5,\mathit{MAXR}{*}20{+}5,1) \ruleimp \mathit{col}(C),\mathit{maxR}(\mathit{MAXR}).\label{ex:2:17} \\[-.7ex]
&&  \mathit{visline}(h(0),5,5,\mathit{MAXC}*20+5,5,1) \ruleimp \mathit{maxC}(\mathit{MAXC}).\label{ex:2:18}\\[-.7ex]
&&  \mathit{visline}(h(R),5,R*20+5,\mathit{MAXC}*20+5,R*20+5,1) \ruleimp \mathit{row}(R),
% \\[-.7ex]
% && \hphantom{\mathit{visline}(h(R),5,R*20+5,\mathit{MAXC}*20+5,R*20+5,1) \ruleimp}
\mathit{maxC}(\mathit{MAXC}).\label{ex:2:19}\\%[1ex]
& \mbox{\tt \%} & \mbox{\tt % Rules 20-24: d
Define possible grid values for editing.}\nonumber\\[-.7ex]
&&  \mathit{vispossiblegridvalues}(\mathit{maze},\mathit{wall}).\label{ex:2:20} \\[-.7ex]
&&  \mathit{vispossiblegridvalues}(\mathit{maze},\mathit{empty}).\label{ex:2:21} \\[-.7ex]
&&  \mathit{vispossiblegridvalues}(\mathit{maze},\mathit{entrance}).\label{ex:2:22} \\[-.7ex]
&&  \mathit{vispossiblegridvalues}(\mathit{maze},\mathit{exit}).\label{ex:2:23}
\end{eqnarray}
\hrule
\caption{Visualisation program for Example~\ref{ex-maze}.}\label{maze-vis-program}
\end{figure}

\subsection{Integration in \sealion}
\kara is written in Java and integrated in the Eclipse-plugin \sealion~\cite{sealion}
for developing answer-set programs.
Currently, it can be used with answer-set programs in the languages of \gringo and \dlv.
\sealion offers functionality to execute external ASP solvers on answer-set programs.
The resulting answer sets can be parsed by the IDE and displayed as expandable
tree structures in a dedicated Eclipse view for interpretations.
Starting from there, the user can invoke \kara
by choosing a pop-up menu entry of the interpretation he or she wants to visualise.
A run configuration dialog will open that allows for choosing
the visualisation program and for setting the solver configuring to be used by
\kara.
Then, the visual editor opens with the generated visualisation.
The process for abducing an interpretation that reflects the modifications to the visualisation
can be started from the visual editor's pop-up menu.
If a respective interpretation exists, one will be added to \sealion's interpretation view.

The sources of \kara and the alpha version of \sealion can be downloaded from
\begin{center}
\url{http://sourceforge.net/projects/mmdasp/}.
\end{center}
An Eclipse update site will be made available as soon as \sealion reaches beta status.

\section{Examples}\label{sec:example}

\begin{figure}[t]
\centering
 \includegraphics[scale=0.6]{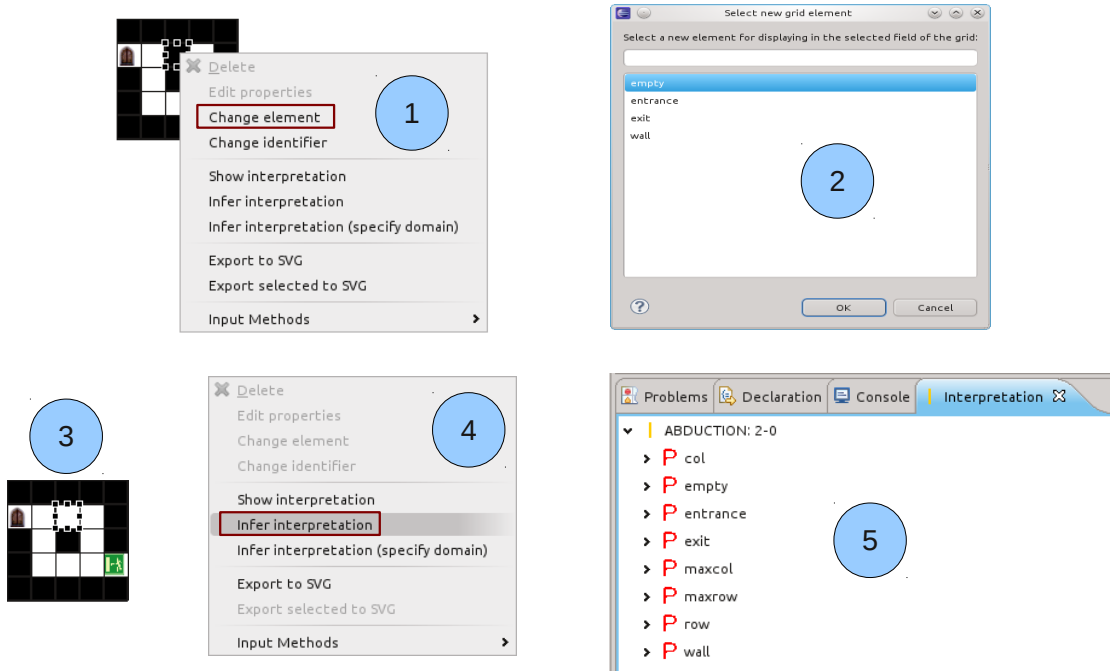}
\caption{Abduction steps in the plugin.}
\label{mazeabd}
\end{figure}

In this section, we provide examples that give an overview of \kara's functionality.
We first illustrate the use of logic grids and the visual editing feature.
\begin{example}\label{ex-maze}
\emph{Maze-generation} is a benchmark problem from the second ASP competition~\cite{competition09}.
The task is to generate a two-dimensional grid, where each cell is either a wall or empty,
that satisfies certain constraints. There are two dedicated empty cells, being the maze's entrance and its exit, respectively.
The following facts represent a sample answer set of a maze generation encoding restricted to interesting predicates.
\[
\begin{array}{l}
\mathit{col}(1..5). \ \mathit{row}(1..5). \ \mathit{maxC}(5). \ \mathit{maxR}(5). \ \mathit{wall}(1,1). \ \mathit{empty}(1,2). \
\mathit{wall}(1,3).\\
 \mathit{wall}(1,4). \ \mathit{wall}(1,5). \ \mathit{wall}(2,1). \ \mathit{empty}(2,2). \
\mathit{empty}(2,3). \ \mathit{empty}(2,4). \ \mathit{wall}(2,5). \\ \mathit{wall}(3,1). \ \mathit{wall}(3,2). \
\mathit{wall}(3,3). \ \mathit{empty}(3,4). \ \mathit{wall}(3,5). \ \mathit{wall}(4,1). \ \mathit{empty}(4,2). \\
\mathit{empty}(4,3). \ \mathit{empty}(4,4). \ \mathit{wall}(4,5). \ \mathit{wall}(5,1). \mathit{wall}(5,2). \
\mathit{wall}(5,3). \ \mathit{empty}(5,4). \\ \mathit{wall}(5,5). \ \mathit{entrance}(1,2). \ \mathit{exit}(5,4). 
\end{array}
\]

Predicates $\mathit{col}/1$ and $\mathit{row}/1$ define indices for the rows and columns of the maze,
while $\mathit{maxC}/1$ and $\mathit{maxR}/1$ give the maximum column and  row number, respectively.
The predicates $\mathit{wall}/2$, $\mathit{empty}/2$, $\mathit{entrance}/2$, and $\mathit{exit}/2$
determine the positions of walls, empty cells, the entrance, and the exit in the grid, respectively.
One may use the visualisation program from Fig.~\ref{maze-vis-program} for maze-generation interpretations of this kind.
%
%%%%%%%%%%%%%%%%%%%%%%%%%%%%%%%%%%%%%%%%%
\begin{figure}[t]
\centering
 \includegraphics[scale=1]{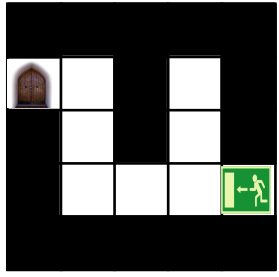}
\caption{Visualisation output for the maze-generation program.}
\label{mazeimg}
\end{figure}
%%%%%%%%%%%%%%%%%%%%%%%%%%%%%%%%%%%%%%%%%

In Fig.~\ref{maze-vis-program}, Rule~(\ref{ex:2:1}) defines a logic grid with identifier $\mathit{maze}$,  $\mathit{MAXR}$ rows, and $\mathit{MAXC}$ columns. The fourth and fifth parameter define the height and width of the grid in pixel. Rule~(\ref{ex:2:2}) is a fact that defines a fixed position for the maze.
The next step is to define the graphical objects to be displayed in the grid.
Because these objects are fixed (i.e., they are used more than once), they can be defined as facts.
A wall is represented by a rectangle with black background and foreground colour\footnote{Black foreground colour is default and may not be set explicitly.} (Rules~(\ref{ex:2:3}) and (\ref{ex:2:4})) whereas an empty cell is rendered as a rectangle with white background and foreground colour (Rules~(\ref{ex:2:5}) to~(\ref{ex:2:7})). The entrance and the exit are represented by two images (Rules~(\ref{ex:2:8}) to~(\ref{ex:2:11})).
Then, these graphical elements are assigned to the respective cell of the grid (Rules~(\ref{ex:2:12}) to (\ref{ex:2:15})).
Rules~(\ref{ex:2:16}) to (\ref{ex:2:19}) render vertical and horizontal lines to better distinguish between the different cells.
Rules~(\ref{ex:2:20}) to (\ref{ex:2:23}) are not needed for visualisation but define possible values for the grid
that we want to be available in the visual editor.

Once the grid is rendered, the user can replace the value of a cell with a value defined using predicate
$\mathit{vispossiblegridvalues}/2$ (e.g., replacing an empty cell with a wall).
The visualisation of the sample interpretation using this program is given in Fig.~\ref{mazeimg}.
Note that the visual representation of the answer set is much easier to cope with than the textual representation of the answer set given in the beginning of the example.

Next, we demonstrate how to use the visual editing feature of \kara to obtain a modified interpretation,
as shown in Fig.~\ref{mazeabd}.
Suppose we want to change the cell $(3,2)$ from being a wall to an empty cell.
The user can select the respective cell and open a pop-up menu that provides an item for changing grid-values.
A dialog opens that allows for choosing among the values that have been defined in the visualisation program,
using the $\mathit{vispossiblegridvalues}/2$ predicate.
When the user has finished editing the visualisation,
he or she can start the abduction process for inferring the new interpretation.
When an interpretation is successfully derived, it is added to \sealion's interpretation view.
\qed\end{example}

%%%%%%%%%%%%%%%%%%%%%%%%%%%%%%%%%%%%%%%%%%%
%%%%%%%%%%%%%%%%%%%%%%%%%%%%%%%%%%%%%%%%%%%%%
{
\kara supports absolute and relative positioning of graphical elements.
If for any visualisation element the predicate $\mathit{visposition}/4$ is defined,
then we have fixed positioning.
Otherwise, the element is positioned automatically.
Then, by default, the elements are randomly positioned on the graphical editor.
However, the user can define the position of an element
\emph{relative} to another element.
This is done by using the predicates
$\mathit{visleft}/2$, $\mathit{visright}/2$, $\mathit{visabove}/2$, $\mathit{visbelow}/2$, and $\mathit{visinfrontof}/2$.
\begin{example}\label{ex:layout}
The following visualisation program makes use of relative positioning for sorting elements according to their label.
\begin{eqnarray}%{rl}
% \mbox{\em \%} & \mbox{\em Rules 1-3.}\\
&&  \mathit{visrect}(a,50,50).\label{eq:ex3:1} \\[-.7ex]
&& \mathit{vislabel}(a,\mathit{laba}).\label{eq:ex3:2} \\[-.7ex]
&& \mathit{vistext}(\mathit{laba},3).\label{eq:ex3:3} \\
% &\mbox{\em \%} & \mbox{\em Rules 4-8.}\\
 && \mathit{vispolygon}(b,0,20,1).\label{eq:ex3:4} \\[-.7ex]
 && \mathit{vispolygon}(b,25,0,2).\label{eq:ex3:5} \\[-.7ex]
 && \mathit{vispolygon}(b,50,20,3).\label{eq:ex3:6} \\[-.7ex]
&& \mathit{vislabel}(b,\mathit{labb}).\label{eq:ex3:7} \\[-.7ex]
 && \mathit{vistext}(\mathit{labb},10).\label{eq:ex3:8} \\
% &\mbox{\em \%} & \mbox{\em Rules 9-11.}\\
&&  \mathit{visellipse}(c,30,30).\label{eq:ex3:9} \\[-.7ex]
 && \mathit{vislabel}(c,\mathit{labc}).\label{eq:ex3:10} \\[-.7ex]
 && \mathit{vistext}(\mathit{labc},5).\label{eq:ex3:11} \\
% \mbox{\em \%} & \mbox{\em Rules 12-14.}\\
&&  \mathit{element}(X) \ruleimp \mathit{visrect}(X,\_,\_).\label{eq:ex3:12} \\[-.7ex]
&& \mathit{element}(X) \ruleimp \mathit{vispolygon}(X,\_,\_,\_).\label{eq:ex3:13} \\[-.7ex]
&&  \mathit{element}(X) \ruleimp \mathit{visellipse}(X,\_,\_).\mathit{element}\hphantom{\ruleimp \mathit{visellipse}(X,\_,\_)}\label{eq:ex3:14} %\\
%
% \end{array}
% \]
% \[
% \begin{array}{rl}
% &\mbox{\em \%} & \mbox{\em Rules 15-16.}\\
\end{eqnarray}
\begin{eqnarray}
&&  \mathit{visleft}(X,Y) \ruleimp \mathit{element}(X),\mathit{element}(Y), \mathit{vislabel}(X,\mathit{LABX}),\nonumber\\[-.7ex]
&& \hphantom{\mathit{visleft}(X,Y) \ruleimp }                     \mathit{vistext}(\mathit{LABX},\mathit{XNUM}), \mathit{vislabel}(Y,\mathit{LABY}),\label{eq:ex3:15} \\[-.7ex]
&& \hphantom{\mathit{visleft}(X,Y) \ruleimp }
                          \mathit{vistext}(\mathit{LABY},\mathit{YNUM}), \mathit{XNUM} < \mathit{YNUM}.\nonumber
\end{eqnarray}

The program defines three graphical objects, a rectangle, a polygon, and an ellipse.
In Rules~(\ref{eq:ex3:1}) to~(\ref{eq:ex3:3}), the rectangle together with its label $3$ is generated.
The shape of the polygon (Rules~(\ref{eq:ex3:4}) to (\ref{eq:ex3:6})) is defined by a sequence of points
relative to the polygon's own coordinate system using the $\mathit{vispolygon}/4$ predicate.
The order in which these points are connected with each other is given by the predicate's fourth argument.
Rules~(\ref{eq:ex3:7}) and (\ref{eq:ex3:8}) generate the label for the polygon and specify its text.
Rules~(\ref{eq:ex3:12}) to (\ref{eq:ex3:14}) state that every rectangle, polygon, and ellipse is an element.
The relative position of the three elements is determined by Rule~(\ref{eq:ex3:15}).
For two elements $E_1$ and $E_2$, $E_1$ has to appear to the left of $E_2$ whenever
the  label of $E_1$ is  smaller than the one of $E_1$.
The output of this visualisation program is given in Fig.~\ref{layoutingimg}.
Note that the visualisation program does not make reference to predicates from an interpretation to visualise,
hence the example illustrates that \kara can also be used for creating arbitrary graphics.
\qed\end{example}
}

\begin{figure}[t]
\centering
 \includegraphics[height=7.5cm]{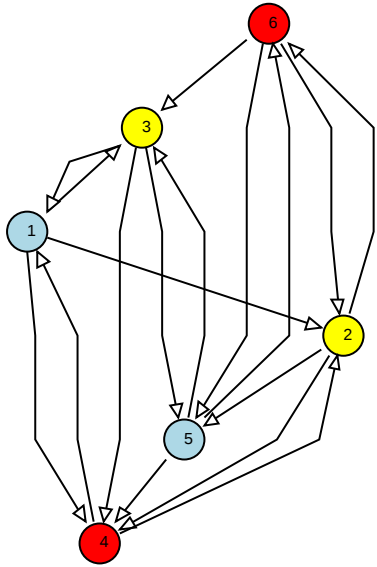}
\caption{Visualisation of a coloured  graph.}
\label{ncoloringimg}
\end{figure}

\begin{figure}[t]
\centering
 \includegraphics[width=14.65cm]{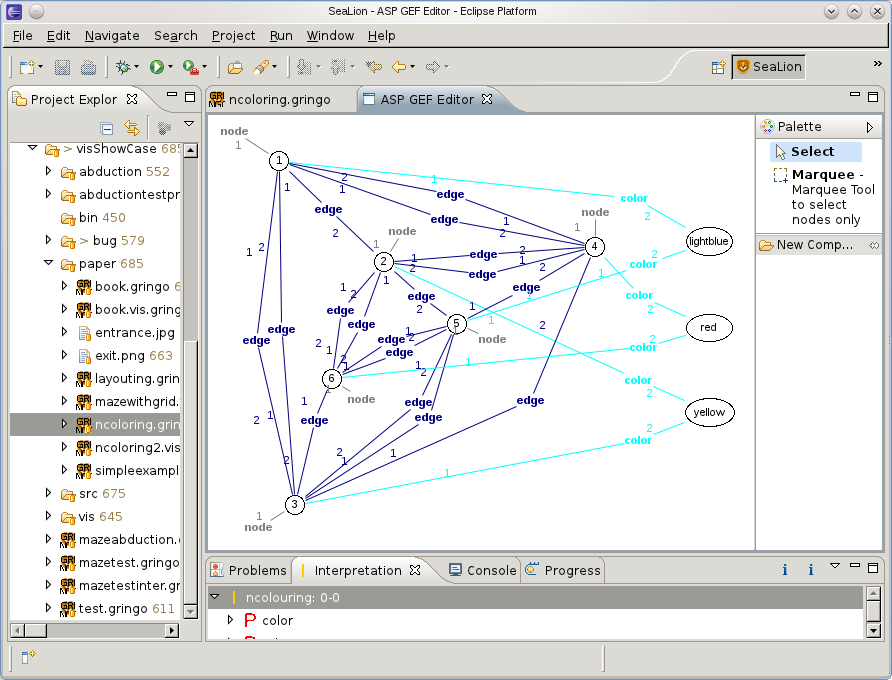}
\caption{A screenshot of \sealion's visual interpretation editor showing a generic visualisation of 
the graph colouring interpretation of Example~\ref{ex:generic} (the layout of the graph has been manually optimised by moving the nodes in the editor).}
\label{fig:autovisimg}
\end{figure}

The last example demonstrates the support for graphs in \kara. Moreover, the generic visualisation feature is illustrated. 
\begin{example}\label{ex:generic}\label{ex:graph}
We want to visualise answer sets of an encoding of a graph-colouring problem.
Assume we have the following interpretation that defines nodes and edges of a graph
as well as a colour for each node.
\[
\begin{array}{r@{}l}
\{&\mathit{node}(1), \ \mathit{node}(2), \ \mathit{node}(3), \ \mathit{node}(4), \ \mathit{node}(5), \ \mathit{node}(6), \
\mathit{edge}(1,2), \ \mathit{edge}(1,3), \\
& \mathit{edge}(1,4), \ \mathit{edge}(2,4), \ \mathit{edge}(2,5), \
\mathit{edge}(2,6), \ \mathit{edge}(3,1), \ \mathit{edge}(3,4), \ \mathit{edge}(3,5), \\
& \mathit{edge}(4,1), \
\mathit{edge}(4,2), \ \mathit{edge}(5,3), \ \mathit{edge}(5,4), \ \mathit{edge}(5,6), \ \mathit{edge}(6,2), \
\mathit{edge}(6,3), \\
&\mathit{edge}(6,5), \ \mathit{color}(1,\mathit{lightblue}), \ \mathit{color}(2,\mathit{yellow}), \
\mathit{color}(3,\mathit{yellow}), \ \mathit{color}(4,\mathit{red}), \\ &\mathit{color}(5,\mathit{lightblue}), \ \mathit{color}(6,\mathit{red})\}. 
\end{array}
\]

\begin{figure}[t]
\centering
 \includegraphics[scale=1.]{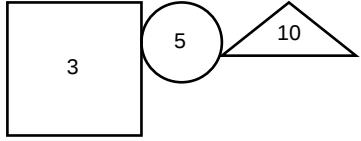}
\caption{Output of the visualisation program in Example~\ref{ex:layout}.}
\label{layoutingimg}
\end{figure}

We make use of the following visualisation program:

% \medskip
% \(
\begin{eqnarray}%{rrl}
&\mbox{\tt \%} & \mbox{\tt %Rule 1. 
Generate a graph.}\nonumber\\[-.7ex]
&&  \mathit{visgraph}(g).\label{eq:ex4:1} \\
&\mbox{\tt \%} & \mbox{\tt % Rules 2-3. 
Generate the nodes of the graph.}\nonumber\\[-.7ex]
&&  \mathit{visellipse}(X,20,20) \ruleimp \mathit{node}(X).\label{eq:ex4:2} \\[-.7ex]
&& \mathit{visisnode}(X,g) \ruleimp \mathit{node}(X).\label{eq:ex4:3} \\
&\mbox{\tt \%} & \mbox{\tt % Rule 4-5. 
Connect the nodes (edges of the input).}\nonumber\\[-.7ex]
&&  \mathit{visconnect}(f(X,Y),X,Y) \ruleimp \mathit{edge}(X,Y).\label{eq:ex4:4} \\[-.7ex]
&&  \mathit{vistargetdeco}(X,\mathit{arrow}) \ruleimp \mathit{visconnect}(X,\_,\_). \label{eq:ex4:5} \\
% \end{array}
% \)
% 
% \[
% \begin{array}{rl}
& \mbox{\tt \%}& \mbox{\tt % Rules 6-8. 
Generate labels for the nodes.}\nonumber\\[-.7ex]
&&  \mathit{vislabel}(X,l(X)) \ruleimp \mathit{node}(X).\label{eq:ex4:6} \\[-.7ex] 
&& \mathit{vistext}(l(X),X) \ruleimp \mathit{node}(X).\label{eq:ex4:7} \\[-.7ex]
&&  \mathit{visfontstyle}(l(X),\mathit{bold}) \ruleimp \mathit{node}(X).\label{eq:ex4:8} \\
& \mbox{\tt \%} &\mbox{\tt % Rule 9. 
Color the node according to the solution.}\nonumber\\[-.7ex]
&&  \mathit{visbackgroundcolor}(X,\mathit{COLOR}) \ruleimp \mathit{node}(X),\mathit{color}(X,\mathit{COLOR}).\label{eq:ex4:9}
\end{eqnarray}
% \]
%
In Rule~(\ref{eq:ex4:1}), a graph, $g$, is defined and a circle for every node from the input interpretation is created (Rule~(\ref{eq:ex4:2})).
Rule~(\ref{eq:ex4:3}) states that each of these circles is logically considered a node of graph $g$.
This has the effect that they will be considered by the algorithm layouting the graph 
during the creation of the visualisation.
The edges of the graph are defined using the $\mathit{visconnect}/3$ predicate (Rule~(\ref{eq:ex4:4})).
It can be used to connect arbitrary graphical elements with a line,
also if they are not nodes of some graph.
As we deal with a directed graph, an arrow is set as target decoration for all the connections (Rule~(\ref{eq:ex4:5})).
Labels for the nodes are set in Rules~(\ref{eq:ex4:6}) to (\ref{eq:ex4:8}).
Finally, Rule~(\ref{eq:ex4:9}) sets the colour of the node according to the interpretation.
The resulting visualisation is depicted in Fig.~\ref{ncoloringimg}.
Moreover, the generic visualisation of the graph colouring interpretation
is given in Fig.~\ref{fig:autovisimg}.
\qed\end{example}

%%%%%%%%%%%%%%%%%%%%%%%%%%%%%%%%%%%%%%%%%%%%%%%%%%%%%%%%%%%
\section{Related Work}\label{sec:related}
%%%%%%%%%%%%%%%%%%%%%%%%%%%%%%%%%%%%%%%%%%%%%%%%%%%%%%%%%%%

The visualisation feature of \kara follows the
previous systems \aspviz~\cite{CliffeVBP08} and \idpdraw~\cite{idpdraw},
which also use ASP for defining how interpretations should be visualised.\footnote{\idpdraw has been used for visualisation of the benchmark problems of the second and third ASP competition.}
Besides the features beyond visualisation, viz.\ the framework for editing visualisations and the support
for multiple solvers,
there are also differences between \kara and these tools regarding visualisation aspects.

\kara allows to write more high-level specifications for positioning the graphical elements of a visualisation.
While \idpdraw and \aspviz require the use of absolute coordinates,
\kara additionally supports relative positioning and automatic layouting for graph and grid structures.
Note that technically, the former is realised using ASP, by guessing positions of the individual elements and adding respective constraints to ensure the correct layout, while the latter is realised by using a standard graph layouting algorithm
which is part of the Eclipse framework.
In \kara, as well as in \idpdraw, each graphical element has a unique identifier
that can be used, \egc to link elements or to set their properties (\egc colour or font style).
That way, programs can be written in a clear and elegant way
since not all properties of an element have to be specified within a single atom.
Here, \kara exploits that the latest ASP solvers  support  function symbols
that allow for generating new identifiers from terms of the interpretation to visualise.
\idpdraw does not support function symbols. 
Instead, for having compound identifiers, \idpdraw uses 
predicates of variable length (\egc $idp\_polygon(id_1, id_2, ...)$).
A disadvantage of this approach is that some solvers, like \dlv, do not support predicates of variable length.
\aspviz does not support identifiers for graphical objects.

The support for a $z$-axis to determine which object should be drawn over others is available in \kara and \idpdraw but missing in \aspviz.
Both \kara and \aspviz support the export of visualisations as vector graphics in the SVG format,
which is not possible with  \idpdraw.
A feature that is supported by \aspviz and \idpdraw, however,
is  creating animations which is not possible with \kara so far.

\kara and \aspviz are written in Java and depend only on a Java Virtual Machine. 
\idpdraw, on the other hand, is written in C++ and depends on the qt libraries.
Finally, \kara is embedded in an IDE, whereas \aspviz and \idpdraw are stand-alone tools.

A related approach from 
software engineering is the Alloy Analyzer, a tool to support the analysis 
of declarative software models~\cite{jackson06}. Models are formulated in a first-order based specification language. 
 The Alloy Analyzer can find satisfying instances of a model using translations to SAT.
Instances of models are first-order structures that can be  automatically visualised as graphs, where the nodes correspond to atoms from
respective signature declarations in the specification, and the edges correspond to relations between atoms. 
Since the Alloy approach is based on finding models for  declarative specifications, it can be regarded as an instance of ASP in
a broader sense. The visualisation of first-order structures in Alloy is closely related to the generic visualisation mode of \kara where no
dedicated visualisation program is needed.
Alloy supports filtering predicates and arguments away of the graph. We consider to add such a feature in future versions of \kara for getting a clearer generic visualisation.

\section{Conclusion}
We presented the tool \kara for visualising and visual editing of interpretations
in ASP.
It supports generic as well as customised visualisations.
For the latter, a powerful language for defining a visualisation
by means of ASP is provided,
supporting, e.g., automated graph layouting, grids of graphical elements, and relative positioning.
The editing feature is based on abductive reasoning, inferring a new interpretation as hypothesis to explain
a modified visualisation.
In future work, we want to add support for defining input and output signatures for programs in \sealion.
Then, the abduction framework of \kara could be easily extended such that
instead of deriving an interpretation that corresponds to the modified visualisation,
one can derive inputs for a domain program such that one of its answer sets has this visualisation.

\appendix

\newcommand{\bts}{}
\newcommand{\ets}{}

\section{Predefined Visualisation Predicates in \kara%
}\label{app:vispred}
\bts
      \begin{longtable}{|p{5.2cm}|p{9.45cm}|}
	\hline
	\textbf{Atom} & \textbf{Intended meaning} \\ \hline\bts
	$\mathit{visellipse}(\mathit{id}$, $\mathit{height}$,$\mathit{width})$ \ets & \bts Defines an ellipse with specified height and width. \ets \\ \hline\bts
	$\mathit{visrect}(\mathit{id}$,$\mathit{height}$,$\mathit{width})$ \ets & \bts Defines a rectangle with specified height and width. \ets \\ \hline\bts
	$\mathit{vispolygon}(\mathit{id}$,$x$,$y$,$\mathit{ord})$ \ets & \bts Defines a point of a polygon. The ordering defines in which order the defined  points are connected with each other. \ets \\ \hline\bts
	$\mathit{visimage}(\mathit{id}$,$\mathit{path})$ \ets & \bts Defines an image given in the specified file. \ets \\ \hline\bts
	$\mathit{visline}(\mathit{id}$,$x_1$,$y_1$,$x_2$,$y_2$,$z)$ \ets & \bts Defines a line between the points $(x_1, y_1)$ and $(x_2, y_2)$. \ets \\ \hline\bts
	$\mathit{visgrid}(\mathit{id}$,$\mathit{rows}$,$\mathit{cols}$,$\mathit{height}$, $\mathit{width})$ \ets & \bts Defines a grid, with the specified number of rows and columns; $\mathit{height}$ and $\mathit{width}$ determine the size of the grid.\ets \\ \hline\bts
	$\mathit{visgraph}(\mathit{id})$ \ets & \bts Defines a graph. \ets \\ \hline\bts 	
	$\mathit{vistext}(\mathit{id}$,$\mathit{text})$ \ets & \bts Defines a text element. \ets \\ \hline\bts
	$\mathit{vislabel}(\mathit{id_g}$,$ \mathit{id_t})$ \ets & \bts Sets the text element $\mathit{id_t}$ as a label for graphical element $\mathit{id_g}$. Labels are supported for the following elements: $\mathit{visellipse}/3$, $\mathit{visrect}/3$, $\mathit{vispolygon}/4$, and $\mathit{visconnect}/3$. \ets \\ \hline\bts
	$\mathit{visisnode}(\mathit{id_n}$,$ \mathit{id_g})$ \ets & \bts Adds the graphical element $\mathit{id_n}$ as a node to a graph $\mathit{id_g}$ for automatic layouting. The following elements are supported as nodes: $\mathit{visrect}/3$, $\mathit{visellipse}/3$, $\mathit{vispolygon}/4$, $\mathit{visimage}/2$. \ets \\ \hline\bts
	$\mathit{visscale}(\mathit{id}$,$ \mathit{height}, \mathit{weight})$ \ets & \bts Scales an image to the specified height and width.\ets \\ \hline\bts
	$\mathit{visposition}(\mathit{id}$,$ x$,$ y$,$ z)$ \ets & \bts Puts an element $\mathit{id}$ on the fixed position $(x,y,z)$. \ets \\ \hline\bts
	$\mathit{visfontfamily}(\mathit{id}$,$ \mathit{ff})$ \ets & \bts Sets the specified font $\mathit{ff}$ for a text element $\mathit{id}$. \ets \\ \hline\bts
	$\mathit{visfontsize}(\mathit{id}$,$ \mathit{size})$ \ets & \bts Sets the font size $\mathit{size}$ for a text element $\mathit{id}$. \ets \\ \hline\bts
	$\mathit{visfontstyle}(\mathit{id}$,$ \mathit{style})$ \ets & \bts Sets the font style for a text element $\mathit{id}$ to bold or italics. \ets \\ \hline\bts
	$\mathit{viscolor}(\mathit{id}$,$ \mathit{color})$ \ets & \bts Sets the foreground colour for the element $\mathit{id}$. \ets \\ \hline\bts
	$\mathit{visbackgroundcolor}(\mathit{id}$,$ \mathit{color})$ \ets & \bts Sets the background colour for the element $\mathit{id}$.
\ets \\ \hline\bts
	$\mathit{visfillgrid}(\mathit{id_g}$,$ \mathit{id_{c}}$,$ \mathit{row}$,$ \mathit{col})$ \ets & \bts Puts element $\mathit{id_{c}}$ in cell $(\mathit{row},\mathit{col})$ of the grid $\mathit{id_g}$. \ets \\ \hline\bts
	$\mathit{visconnect}(\mathit{id_c}$,$ \mathit{id_{g_1}}$,$ \mathit{id_{g_2}})$ \ets & \bts Connects two elements, $\mathit{id_{g_{1}}}$ and $\mathit{id_{g_{2}}}$, by a line such that  $\mathit{id_{g_{1}}}$ is the source and $\mathit{id_{g_{2}}}$ is the target of the connection. \ets \\ \hline\bts
	$\mathit{vissourcedeco}(\mathit{id}$,$ \mathit{deco})$ \ets & \bts Sets the source decoration for a connection. \ets \\ \hline\bts
	$\mathit{vistargetdeco}(\mathit{id}$,$ \mathit{deco})$ \ets & \bts Sets the target decoration for a connection.\ets \\ \hline\bts
	$\mathit{visleft}(\mathit{id_l}$,$ \mathit{id_r})$ \ets & \bts Ensures that the $x$-coordinate of $\mathit{id_l}$ is less than that of $\mathit{id_r}$. \ets \\ \hline\bts
	$\mathit{visright}(\mathit{id_r}$,$ \mathit{id_l})$ \ets & \bts Ensures that the $x$-coordinate of $\mathit{id_r}$ is greater than that of $\mathit{id_l}$. \ets \\ \hline\bts
	$\mathit{visabove}(\mathit{id_t}$,$ \mathit{id_b})$ \ets & \bts Ensures that the $y$-coordinate of $\mathit{id_t}$ is smaller than that of $\mathit{id_b}$. \ets \\ \hline\bts
	$\mathit{visbelow}(\mathit{id_b}$,$ \mathit{id_t})$ \ets & \bts Ensures that the $y$-coordinate of $\mathit{id_b}$ is greater than that of $\mathit{id_t}$. \ets \\ \hline\bts
	$\mathit{visinfrontof}(\mathit{id_1}$,$ \mathit{id_2})$ \ets & \bts Ensures that the $z$-coordinate of $\mathit{id_1}$ is greater than that of $\mathit{id_2}$. \ets \\ \hline\bts
	$\mathit{vishide}(\mathit{id})$ \ets & \bts Hides the element $\mathit{id}$. \ets \\ \hline\bts
	$\mathit{visdeletable}(\mathit{id})$ \ets & \bts Defines that the element $\mathit{id}$ can be deleted in the visual editor.\ets \\ \hline\bts
	$\mathit{viscreatable}(\mathit{id})$ \ets & \bts Defines that the element $\mathit{id}$ can be created in the visual editor. \ets \\ \hline\bts
	$\mathit{vischangable}(\mathit{id}$,$ \mathit{prop})$ \ets & \bts Defines that the property $\mathit{prop}$ can be changed for the element $\mathit{id}$ in the visual editor.\ets \\ \hline\bts
	$\mathit{vispossiblegridvalues}(\mathit{id}$,$ \mathit{id_e})$ \ets & \bts Defines that graphical element $\mathit{id_e}$ is available as possible grid value for a grid $\mathit{id}$ in the visual editor. \ets \\ \hline
      \end{longtable}
\ets

\end{document}